\documentclass[prd,reprint,showpacs,showkeys]{revtex4-1}
\usepackage{amsfonts}
\usepackage{amssymb}
\usepackage{amsmath}
\usepackage{graphicx}
\usepackage[font={footnotesize,it}]{caption}

\input{tcilatex}

\begin{document}

\title{Effect of the Gauss-Bonnet parameter in the stability of thin-shell
wormholes}
\author{Z. Amirabi}
\email{zahra.amirabi@emu.edu.tr}
\author{M. Halilsoy}
\email{mustafa.halilsoy@emu.edu.tr}
\author{S. Habib Mazharimousavi}
\email{habib.mazhari@emu.edu.tr}
\affiliation{Department of Physics, Eastern Mediterranean University, G. Magusa, north
Cyprus, Mersin 10, Turkey. }
\date{\today }

\begin{abstract}
We study the stability of thin-shell wormholes in
Einstein-Maxwell-Gauss-Bonnet gravity. The equation of state of the thin
shell wormhole is considered first to obey a generalized Chaplygin gas and
then we generalize it to an arbitrary state function which covers all known
cases studied so far. In particular we study the modified Chaplygin gas and
give an assessment for a general parotropic fluid. Our study is in $d-$%
dimensions and with numerical analysis in $d=5$ we show the effect of the GB
parameter in the stability of thin-shell wormholes against the radial
perturbations. \qquad
\end{abstract}

\maketitle

\section{Introduction}

In an attempt to minimize the exotic matter of a traversable wormhole, Matt
Visser introduced the concept of thin-shell wormhole (TSW) \cite{1}. More
precisely, in \cite{2} two copies of the Schwarzschild spacetimes are cut
and glued to make the TSW. On the other hand Brady, Louko and Poisson
studied the stability of a thin shell around a black hole in \cite{3}. In
that work, using the Israel's junction conditions \cite{4}, the mechanical
stability of a static, spherically symmetric massive thin shell has been
investigated. Following this work Poisson and Visser in \cite{5} considered
the stability of the TSW against linearized perturbations around some static
spherically symmetric solutions of the Einstein equations. In that paper, in
particular, the form of equation of state of the matter which supports the
TSW was chosen to be $p=p\left( \sigma \right) $ and following the
calculation a parameter $\beta ^{2}\left( \sigma \right) \equiv \frac{%
\partial p}{\partial \sigma }$ has been defined which plays important role
for having a stable TSW. Irrespective of the form of $p\left( \sigma \right)
,$ it was shown that $\frac{\partial p}{\partial \sigma }$ at the static
configuration which occurs at $a=a_{0}$ the equilibrium radius of the throat
of the TSW appears in the final condition. The idea of TSW and its stability
have been developed and generalized in many directions. Ishak and Lake, in
their work \cite{6} has continued along the previous line by adding the
cosmological constant into the solution of the bulk spacetime. Eiroa and
Simeone \cite{7} developed the cylindrical TSW, Lobo studied the phantom
wormholes and their stability in \cite{8}, while TSW in dilaton gravity has
been introduced in \cite{9}. A generic, dynamic spherically symmetric
thin-shell and its corresponding stability has been discussed in \cite{10}.
Chaplygin gas traversable wormholes and generalized Chaplygin gas supported
spherically symmetric TSW have been discussed in \cite{11} while higher
dimensional static spherically symmetric TSW in Einstein-Maxwell theory was
studied by Rahaman, Kalam and Chakraborty in \cite{12}. Vacuum thin shell
solutions in five-dimensional Lovelock gravity has been studied in \cite{SW}%
. Extension toward the Einstein-Maxwell-Gauss-Bonnet (EMGB) was investigated
in \cite{13} and its stability and existence of TSW supported by normal
matter in \cite{14}  .The non-asymptotically flat TSW in higher dimensional
spherically symmetric Einstein-Yang-Mills theory has been considered in \cite%
{15} and its extension to Einstein-Yang-Mills-Gauss-Bonnet is given in \cite%
{16}. TSW in Ho\v{r}ava-Lifshitz gravity was introduced in \cite{17} and TSW
in Lovelock modified theory of gravity has been given in \cite{18}. In \cite%
{19}, rotating TSW in Kerr spacetime was found and TSW in Brans-Dicke theory
and its stability were investigated in \cite{20}. Furthermore, TSW in Dvali,
Gabadadze and Porrati (DGP) theory is determined in \cite{21} while the TSW
in Einstein-nonlinear Maxwell theory has been found in \cite{22}.

The above list is not complete and there are some other works which in some
senses generalized the idea of TSW introduced in \cite{1,2}. Another form of
generalization also is going on parallel to the concept of TSW which is the
Israel junction conditions \cite{4}. In \cite{23} the generalized
Darmois-Israel boundary conditions has been worked out and using it
generalized junction conditions in Einstein-Gauss-Bonnet (EGB) gravity and
in third order Lovelock gravity have been found in \cite{16,18}. For the
whole set of Lovelock theories, the Israeal junction conditions have been
generalized by Gravanisa and Willison in \cite{24}.

Among other aspects the foremost challenging problems related to TSW \cite%
{1,2,3,4,5,6,7,8,9,10,11,12,13,14,15,16,17,18,19,20,21,22} are, $i$)
positivity of energy density, and $ii$) stability against symmetry
preserving perturbations. To overcome these problems recently there have
been various attempts in EGB gravity with Maxwell and Yang-Mills sources.
Specifically, with the negative Gauss-Bonnet (GB) parameter $\left( \alpha
<0\right) $ we obtained stable TSW, obeying a linear equation of state,
against radial perturbations \cite{14}. By linear equation of state it is
meant that the energy density $\sigma $ and surface pressure $p$ satisfy a
linear relation. To respond the other challenge, however, i.e. the
positivity of the energy density $\left( \sigma >0\right) $, we maintain
still a cautious optimism. To be realistic, only in the case of
Einstein-Yang-Mills-Gauss-Bonnet (EYMGB) theory and in a finely-tuned narrow
band of parameters we were able to beat both of the above stated challenges 
\cite{14}. Our stability analysis with the negative energy density was
extended further to cover non-asymptotically flat (NAF) dilatonic solutions 
\cite{15}.

In this paper we show that stability analysis of TSW extends to the case of
a generalized Chaplygin gas (GCG) which has already been considered within
the context of Einstein-Maxwell TSWs \cite{4}. Due to the accelerated
expansion of our universe a repulsive effect of a Chaplygin gas (CG) has
been considered widely in recent times. By the same token therefore \ it
would be interesting to see how a GCG supports a TSW against radial
perturbations in GB gravity. For this purpose we perturb the TSW radially
and reduce the equation into a particle in a potential well problem with
zero total energy. The stability amounts to the determination of the
positive domain for the second derivative of the potential. We obtain plots
that provides us such physical regions indicating stable wormholes. Beside
the example of a GCG we consider an equation of state with quite generality.
Namely, the relation between the pressure $p$ and the energy density $\sigma 
$ is given by the parotropic form $p=\psi \left( \sigma \right) $, for an
arbitrary function $\psi \left( \sigma \right) $. The stability criteria for
such a wormhole have been derived as well.

Organization of the paper is as follows. In Sec. II we introduce our
formalism of TSW in EMGB theory. Stability problem of the obtained TSW
supported by GCG is considered in Sec. III. In Sec. IV we generalize our
equation of state further and consider cases other than the GCG. The paper
ends with our Conclusion in Sec. V.

\section{TSW in EMGB gravity}

The $d-$dimensional EMGB action without cosmological constant 
\begin{equation}
S=\frac{1}{16\pi G}\int \sqrt{\left\vert g\right\vert }d^{d}x\left( R+\alpha 
\mathcal{L}_{GB}-\frac{1}{4}\mathcal{F}\right) .
\end{equation}%
where $G$ is the $d-$dimensional Newton constant, $\mathcal{F=}F_{\mu \nu
}F^{\mu \nu }$ is the Maxwell invariant and $\alpha $ is the GB parameter
with Lagrangian 
\begin{equation}
\mathcal{L}_{GB}=R^{2}-4R_{\mu \nu }R^{\mu \nu }+R_{\mu \nu \rho \sigma
}R^{\mu \nu \rho \sigma }.
\end{equation}%
Variation of $S$ with respect to $g_{\mu \nu }$ yields the EMGB field
equations, 
\begin{equation}
G_{\mu \nu }+2\alpha H_{\mu \nu }^{\ }=T_{\mu \nu }
\end{equation}%
in which $H_{\mu \nu }$ and $T_{\mu \nu }$ are given by 
\begin{multline}
H_{\mu \nu }^{\ }=2\left( -R_{\mu \text{ \ \ }}^{\ \sigma \kappa \tau
}R_{\nu \sigma \kappa \tau }-2R_{\ \mu \rho \nu \sigma }^{\quad }R^{\rho
\sigma }-\right. \\
\left. 2R_{\mu \sigma }R_{\text{ \ }\nu }^{\sigma }+RR_{\mu \nu }^{\
}\right) -\frac{1}{2}g_{\mu \nu }^{\ }\tciLaplace _{GB},
\end{multline}

\begin{equation}
T_{\mu \nu }=F_{\mu \alpha }F_{\nu }^{\;\alpha }-\frac{1}{4}g_{\mu \nu
}F_{\alpha \beta }F^{\alpha \beta }.
\end{equation}%
Our static spherically symmetric metric ansatz will be%
\begin{equation}
ds^{2}=-f\left( r\right) dt^{2}+\frac{dr^{2}}{f\left( r\right) }%
+r^{2}d\Omega _{d-2}^{2},
\end{equation}%
in which 
\begin{equation}
d\Omega _{d-2}^{2}=d\theta _{1}^{2}+\underset{i=2}{\overset{d-2}{\dsum }}%
\underset{j=1}{\overset{i-1}{\dprod }}\sin ^{2}\theta _{j}\;d\theta _{i}^{2}
\end{equation}%
\begin{equation*}
0\leq \theta _{d-2}\leq 2\pi ,0\leq \theta _{i}\leq \pi ,1\leq i\leq d-3
\end{equation*}%
and $f\left( r\right) $ is to be found.

Construction of the thin-shell wormhole in the static spherically symmetric
spacetime follows the standard procedure used before \cite{1,2,3}. In this
method we consider two copies $\mathcal{M}_{1,2}$ of the spacetime 
\begin{equation}
\mathcal{M}_{1,2}=\left\{ \left. \left( t,r,\theta _{1},...,\theta
_{d-2}\right) \right\vert r\geq a,\text{ \ }a>r_{h}\right\}
\end{equation}%
which are egotistically incomplete manifolds whose boundaries are given by
the following timelike hypersurface%
\begin{multline}
\Sigma _{1,2\text{ }}= \\
\left\{ \left. \left( t,r,\theta _{1},...,\theta _{d-2}\right) \right\vert
F\left( r\right) =r-a=0,\text{ \ }a>r_{h}\right\} .
\end{multline}%
By identifying the above hypersurfaces on $r=a$ one gets a geodesically
complete manifold $\mathcal{M=M}_{1}\cup \mathcal{M}_{2}.$

We introduce the induced coordinates on the wormhole $\xi ^{a}=\left( \tau
,\theta _{1},\theta _{2},...\right) $ - with $\tau $ the proper time - in
terms of the original bulk coordinates $x^{\gamma }=\left( t,r,\theta
_{1},...,\theta _{d-2}\right) .$ Further to the Israel junction conditions 
\cite{4}, the generalized Darmois-Israel boundary conditions \cite{23}, are
chosen for the case of EMGB modified gravity. The latter conditions on $%
\Sigma $ take the form 
\begin{multline}
2\left\langle K_{ab}-Kh_{ab}\right\rangle + \\
4\alpha \left\langle 3J_{ab}-Jh_{ab}+2P_{acdb}K^{cd}\right\rangle =-\kappa
^{2}S_{ab},
\end{multline}%
in which $\left\langle .\right\rangle $ stands for a jump across the
hypersurface $\Sigma =\Sigma _{1\text{ }}=\Sigma _{2\text{ }}$, $%
h_{ab}=g_{ab}-n_{a}n_{b}$ is the induced metric on $\Sigma $ with normal
vector $n_{a}$ and $S_{a}^{b}=$diag$\left( \sigma ,p_{\theta _{1}},p_{\theta
_{2}},...\right) $ is the energy momentum tensor on the thin-shell. Therein
the extrinsic curvature $K_{ab}^{\pm }$(with trace $K$) is defined as 
\begin{equation}
K_{ab}^{\pm }=-n_{c}^{\pm }\left( \frac{\partial ^{2}x^{c}}{\partial \xi
^{a}\partial \xi ^{b}}+\Gamma _{mn}^{c}\frac{\partial x^{m}}{\partial \xi
^{a}}\frac{\partial x^{n}}{\partial \xi ^{b}}\right) _{r=a}.
\end{equation}%
The divergence-free part of the Riemann tensor $P_{abcd}$ and the tensor $%
J_{ab}$ (with trace $J$) are given also by%
\begin{multline}
P_{abcd}=R_{abcd}+\left( R_{bc}h_{da}-R_{bd}h_{ca}\right) - \\
\left( R_{ac}h_{db}-R_{ad}h_{cb}\right) +\frac{1}{2}R\left(
h_{ac}h_{db}-h_{ad}h_{cb}\right) ,
\end{multline}%
\begin{multline}
J_{ab}= \\
\frac{1}{3}\left[
2KK_{ac}K_{b}^{c}+K_{cd}K^{cd}K_{ab}-2K_{ac}K^{cd}K_{ab}-K^{2}K_{ab}\right] .
\end{multline}

The black hole solution of the EMGB field equations (with $\Lambda =0$) is
given by \cite{25} 
\begin{multline}
f_{\pm }\left( r\right) =1+\frac{r^{2}}{2\tilde{\alpha}}\times  \\
\left( 1\pm \sqrt{1+4\tilde{\alpha}\left( \frac{2M}{8\pi r^{d-1}}-\frac{Q^{2}%
}{2\left( d-2\right) \left( d-3\right) r^{2\left( d-2\right) }}\right) }%
\right) 
\end{multline}%
in which $\tilde{\alpha}=\left( d-3\right) \left( d-4\right) \alpha ,$ $M$
is an integration constant related to the ADM mass of the BH and $Q$ is the
electric charge of the BH. (We must comment that in the rest of the paper we
assume $\alpha \geq 0$ and the calculations are based on the negative branch
solution i.e., $f\left( r\right) =f_{-}\left( r\right) $.) The corresponding
electric field $2-$form is given by%
\begin{equation}
\mathbf{F=}\frac{Q}{r^{2\left( d-2\right) }}dt\wedge dr.
\end{equation}%
The components of energy momentum tensor on the thin shell are%
\begin{multline}
\sigma =-S_{\tau }^{\tau }= \\
-\frac{\Delta \left( d-2\right) }{8\pi }\left[ \frac{2}{a}-\frac{4\tilde{%
\alpha}}{3a^{3}}\left( \Delta ^{2}-3\left( 1+\dot{a}^{2}\right) \right) %
\right] ,
\end{multline}

\begin{multline}
p=S_{\theta _{i}}^{\theta _{i}}=\frac{1}{8\pi }\left\{ \frac{2\left(
d-3\right) \Delta }{a}+\frac{2\ell }{\Delta }\right. - \\
\frac{4\tilde{\alpha}}{3a^{2}}\left[ 3\ell \Delta -\frac{3\ell }{\Delta }%
\left( 1+\dot{a}^{2}\right) +\frac{\Delta ^{3}}{a}\left( d-5\right) \right. -
\\
\left. \left. \frac{6\Delta }{a}\left( a\ddot{a}+\frac{d-5}{2}\left( 1+\dot{a%
}^{2}\right) \right) \right] \right\} ,
\end{multline}%
in which $\ell =\ddot{a}+f_{\pm }^{\prime }\left( a\right) /2$, $\Delta =%
\sqrt{f_{\pm }\left( a\right) +\dot{a}^{2}}$ and while a 'dot' implies
derivative with respect to the proper time $\tau $ a 'prime' denotes
differentiation with respect to the argument of the function. These
expressions pertain to the static configuration if we consider $a=a_{0}=$%
constant and therefore%
\begin{multline}
\sigma _{0}=-\frac{\sqrt{f_{\pm }\left( a_{0}\right) }\left( d-2\right) }{%
8\pi }\times \\
\left[ \frac{2}{a_{0}}-\frac{4\tilde{\alpha}}{3a_{0}^{3}}\left( f_{\pm
}\left( a_{0}\right) -3\right) \right] ,
\end{multline}%
\begin{multline}
p_{0}=\frac{\sqrt{f_{\pm }\left( a_{0}\right) }}{8\pi }\left\{ \frac{2\left(
d-3\right) }{a_{0}}+\frac{f_{\pm }^{\prime }\left( a_{0}\right) }{f_{\pm
}\left( a_{0}\right) }-\frac{4\tilde{\alpha}}{3a_{0}^{2}}\right. \\
\left. \left[ \frac{3}{2}f_{\pm }^{\prime }\left( a_{0}\right) -\frac{%
3f_{\pm }^{\prime }\left( a_{0}\right) }{2f_{\pm }\left( a_{0}\right) }%
+\left( d-5\right) \left( \frac{f_{\pm }\left( a_{0}\right) -3}{a_{0}}%
\right) \right] \right\} .
\end{multline}%
We add also that in the case of a dynamic throat the conservation equation
amounts to%
\begin{equation}
\frac{d}{d\tau }\left( \sigma a^{\left( d-2\right) }\right) +p\frac{d}{d\tau 
}\left( a^{\left( d-2\right) }\right) =0.
\end{equation}

\begin{figure}[tbp]
\includegraphics[width=80mm,scale=0.7]{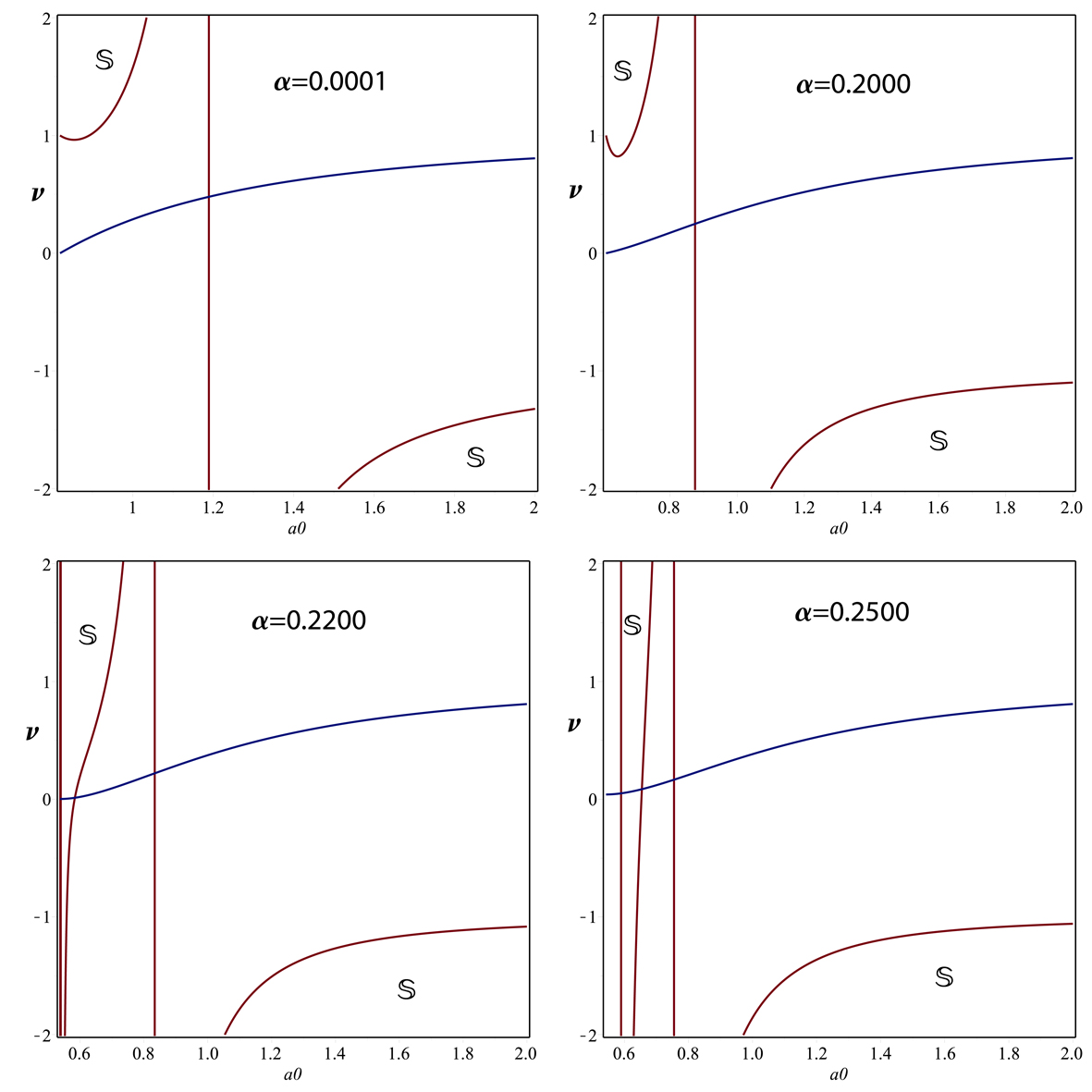}
\caption{Stability region in terms of $\protect\nu $ and radius of the
throat $a_{0}$ for $d=5,$ $M=20$, $Q=1$ and various values of $\protect%
\alpha .$ The stable region is denoted by $\mathrm{S.}$ The metric function
is also displayed for $r$ larger than the horizon.}
\end{figure}

\section{Stability of the EMGBTSW supported by GCG}

Our aim in the sequel is to perturb the throat of the thin-shell wormhole
radially around the equilibrium radius $a_{0}.$ To do this, we assume that
the equation of state is in the form of a GCG \cite{11}, i.e., 
\begin{equation}
p=\left( \frac{\sigma _{0}}{\sigma }\right) ^{\nu }p_{0}
\end{equation}%
in which $\nu \in \left( 0,1\right] $ is a free parameter and $\sigma
_{0}/p_{0}$ correspond to $\sigma /p$ at the equilibrium radius $a_{0}$. We
plug in the latter expression into the conservation energy equation (20) to
find a closed form for the dynamic tension on the thin-shell after
perturbation as follows 
\begin{multline}
\sigma \left( a\right) =\sigma _{0}\left[ \left( \frac{a_{0}}{a}\right)
^{\left( 1+\nu \right) \left( d-2\right) }+\right. \\
\left. \frac{p_{0}}{\sigma _{0}}\left( \left( \frac{a_{0}}{a}\right)
^{\left( 1+\nu \right) \left( d-2\right) }-1\right) \right] ^{\frac{1}{1+\nu 
}}.
\end{multline}%
Equating this with the one found in Eq. (16), one finds a particle-like
equation of motion 
\begin{equation}
\dot{a}^{2}+V\left( a\right) =0,
\end{equation}%
which describes the behavior of the throat after the perturbation. The
intricate potential $V\left( a\right) $ satisfies 
\begin{multline}
\sigma =-\frac{\sqrt{f_{\pm }\left( a\right) -V\left( a\right) }\left(
d-2\right) }{8\pi }\times \\
\left[ \frac{2}{a}-\frac{4\tilde{\alpha}}{3a^{3}}\left( f_{\pm }\left(
a\right) +2V\left( a\right) -3\right) \right]
\end{multline}%
in which $\sigma $ is given by (22). At the static configuration at which $%
a=a_{0}$ one can show that $V\left( a_{0}\right) =0$ and $V^{\prime }\left(
a_{0}\right) =0.$ This implies that Eq. (23) can be expanded about $a=a_{0}$
such that%
\begin{equation}
\dot{x}^{2}+\frac{1}{2}V^{\prime \prime }\left( a_{0}\right) x^{2}=0,
\end{equation}%
in which $x=a-a_{0}.$ Derivative of the latter equation with respect to $%
\tau ,$ yields%
\begin{equation}
\ddot{x}+\frac{1}{2}V^{\prime \prime }\left( a_{0}\right) x=0,
\end{equation}%
which upon $V^{\prime \prime }\left( a_{0}\right) \geq 0$ admits an
oscillatory motion or stability of the thin shell wormhole at $a=a_{0}$. The
exact form of $V^{\prime \prime }\left( a_{0}\right) $ is given by%
\begin{equation}
V^{\prime \prime }\left( a_{0}\right) =\frac{\mathfrak{B}_{1}\nu +\mathfrak{B%
}_{2}}{2a_{0}^{2}f_{0}\left[ 3a_{0}^{2}-2\tilde{\alpha}\left( 3-f_{0}\right) %
\right] \left[ a_{0}^{2}+2\tilde{\alpha}\left( 1+f_{0}\right) \right] }
\end{equation}%
where%
\begin{multline}
\mathfrak{B}_{1}=6\left[ -\frac{2\tilde{\alpha}\left( d-5\right) f_{0}^{2}}{3%
}+\right. \\
\left. \left[ \left( -f_{0}^{\prime }a_{0}+2\left( d-5\right) \right) \tilde{%
\alpha}+a_{0}^{2}\left( d-3\right) \right] f_{0}+\frac{f_{0}^{\prime
}a_{0}\left( a_{0}^{2}+2\tilde{\alpha}\right) }{2}\right] \\
\left[ 4f_{0}^{2}\tilde{\alpha}+\left( -2\tilde{\alpha}f_{0}^{\prime
}a_{0}-2a_{0}^{2}-12\tilde{\alpha}\right) f_{0}+f_{0}^{\prime }a_{0}\left(
a_{0}^{2}+2\tilde{\alpha}\right) \right]
\end{multline}%
and

\begin{multline}
\mathfrak{B}_{2}=-16\tilde{\alpha}^{2}\left( d-5\right) f_{0}^{4}+8\tilde{%
\alpha}f_{0}^{3} \\
\left[ \left( \tilde{\alpha}f_{0}^{\prime \prime }-18+4d\right) a_{0}^{2}+%
\tilde{\alpha}f_{0}^{\prime }\left( d-7\right) a_{0}+12\tilde{\alpha}\left(
d-5\right) \right] \\
+\left\{ \left[ \left( 4f_{0}^{\prime 2}-32f_{0}^{\prime \prime }\right)
a_{0}^{2}-32\left( d-7\right) f_{0}^{\prime }a_{0}\right. \right. - \\
\left. 144\left( d-5\right) \right] \tilde{\alpha}^{2}-16\left[
f_{0}^{\prime \prime }a_{0}^{2}+\left( d-6\right) f_{0}^{\prime
}a_{0}\right. + \\
\left. \left. 6\left( d-4\right) -3\right] a_{0}^{2}\tilde{\alpha}%
-12a_{0}^{4}\tilde{\alpha}\left( d-3\right) \right\} f_{0}^{2}+2 \\
\left[ 3a_{0}^{3}f_{0}^{\prime \prime }+3\left( d-3\right)
a_{0}^{2}f_{0}^{\prime }-\right. \left. 2\tilde{\alpha}\left( f_{0}^{\prime
2}-3f_{0}^{\prime \prime }\right) a_{0}+6\tilde{\alpha}f_{0}^{\prime }\left(
d-7\right) \right] \\
\times \left( a_{0}^{2}+2\tilde{\alpha}\right)
a_{0}f_{0}-3a_{0}^{2}f_{0}^{\prime 2}\left( a_{0}^{2}+2\tilde{\alpha}\right)
^{2}.
\end{multline}

Fig. 1 depicts a $5-$dimensional plot of stable region with respect to $%
a_{0} $ and $\nu $ with $M=20,$ $Q=1$ and variable $\tilde{\alpha}.$ The
stable regions are indicated by letter $\mathrm{S.}$ As it is displayed in
Fig. 1 the stability region has two parts in each case, the area in negative 
$\nu $ and positive $\nu .$ The former is almost for $\nu <-1$ which is not
a physical state. The latter contains partly the interval $\nu \in \left( 0,1%
\right] $ which is in our interest. We observe that by increasing $\tilde{%
\alpha}$ this physical stable region develops and therefore the TSW is more
stable. In addition to the stable regions in Fig. 1, we plot the metric
function to give an estimation of the location of the horizon for the same
parameters.

\begin{figure}[tbp]
\includegraphics[width=70mm,scale=0.7]{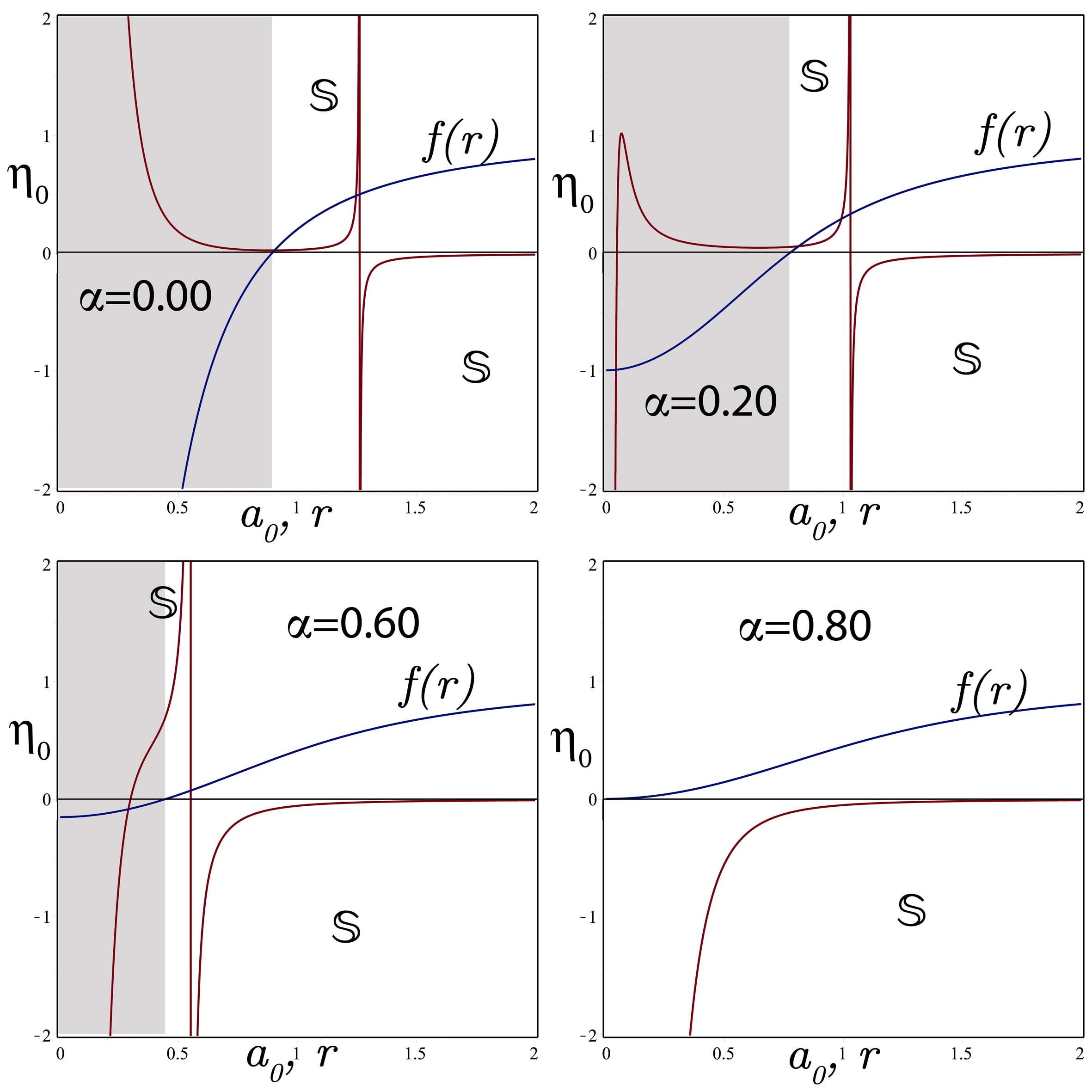}
\caption{Stability region in terms of $\protect\eta _{0}$ and radius of the
throat $a_{0}$ for CG ($\protect\nu =1,\protect\xi _{0}=0$)$,$ $d=5,$ $M=20$%
, $Q=0$ and various values of $\protect\alpha .$ The stable region is
denoted by $\mathrm{S}$ which is identified by $V^{\prime \prime }\left(
a_{0}\right) >0,$ from Eqs. (27-29)$\mathrm{.}$ The metric function is also
displayed in terms of $r$. The shaded region is for $r<r_{h}$ in which $%
r_{h} $ is the event horizon.}
\end{figure}

\begin{figure}[tbp]
\includegraphics[width=70mm,scale=0.7]{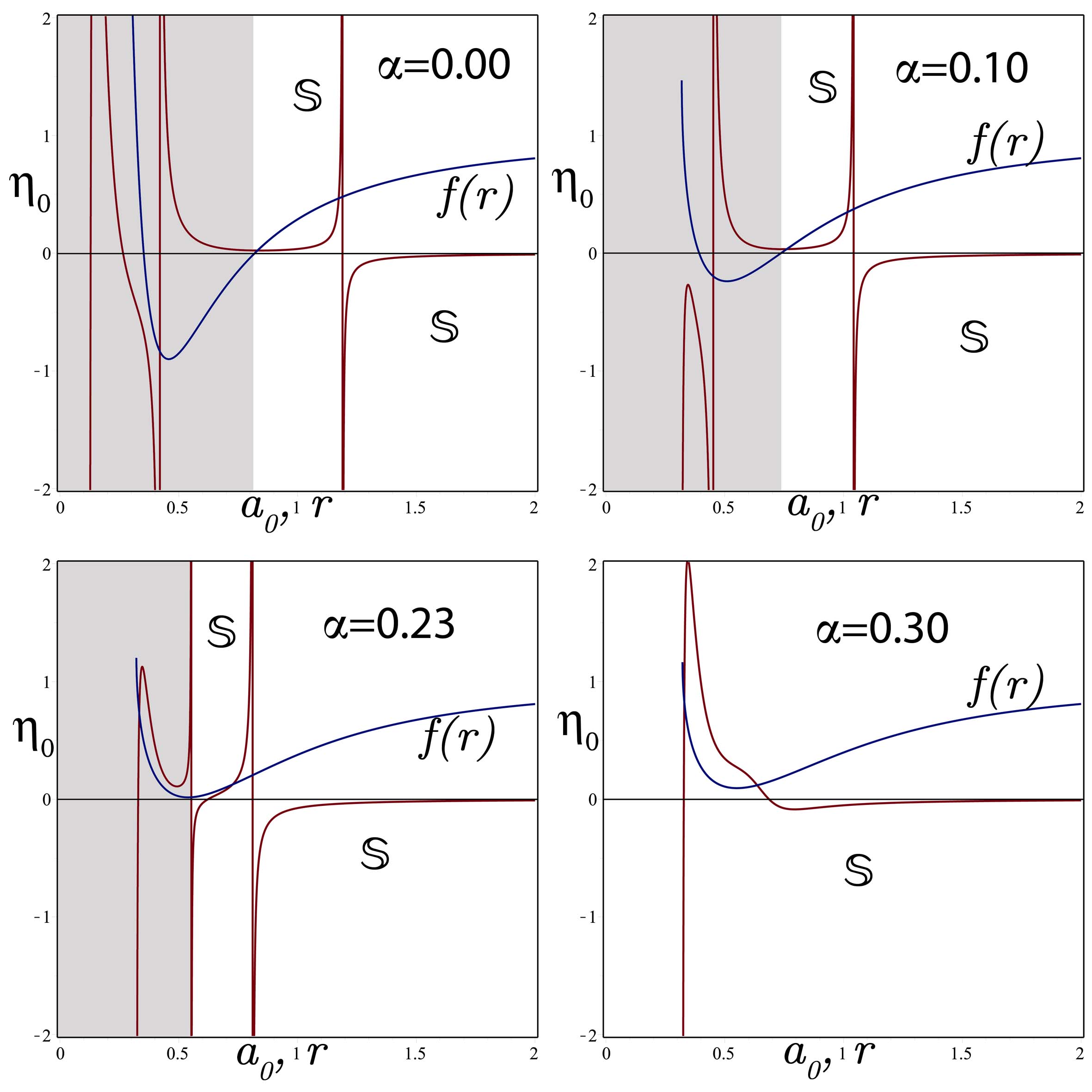}
\caption{Stability region in terms of $\protect\eta _{0}$ and radius of the
throat $a_{0}$ for CG ($\protect\nu =1,\protect\xi _{0}=0$)$,$ $d=5,$ $M=20$%
, $Q=1$ and various values of $\protect\alpha .$ The stable region is
denoted by $\mathrm{S.}$ The metric function is also displayed in terms of $%
r $. The shaded region is for $r<r_{h}$ in which $r_{h}$ is the event
horizon.}
\end{figure}

\begin{figure}[tbp]
\includegraphics[width=70mm,scale=0.7]{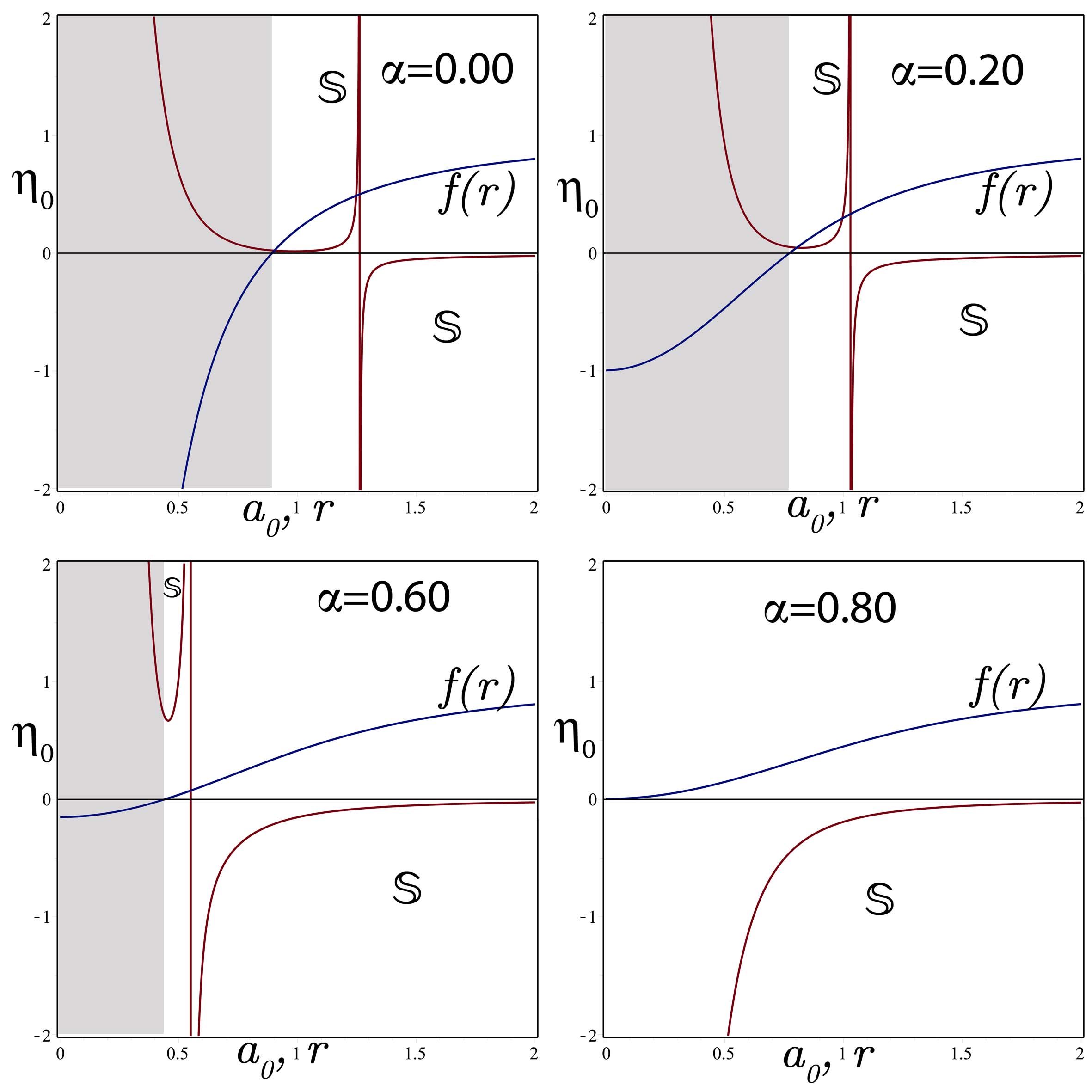}
\caption{Stability region in terms of $\protect\eta _{0}$ and radius of the
throat $a_{0}$ for MCG ($\protect\nu =1,\protect\xi _{0}=1$)$,$ $d=5,$ $M=20$%
, $Q=0$ and various values of $\protect\alpha .$ The stable region is
denoted by $\mathrm{S.}$ The metric function is also displayed in terms of $%
r $. The shaded region is for $r<r_{h}$ in which $r_{h}$ is the event
horizon.}
\end{figure}

\begin{figure}[tbp]
\includegraphics[width=70mm,scale=0.7]{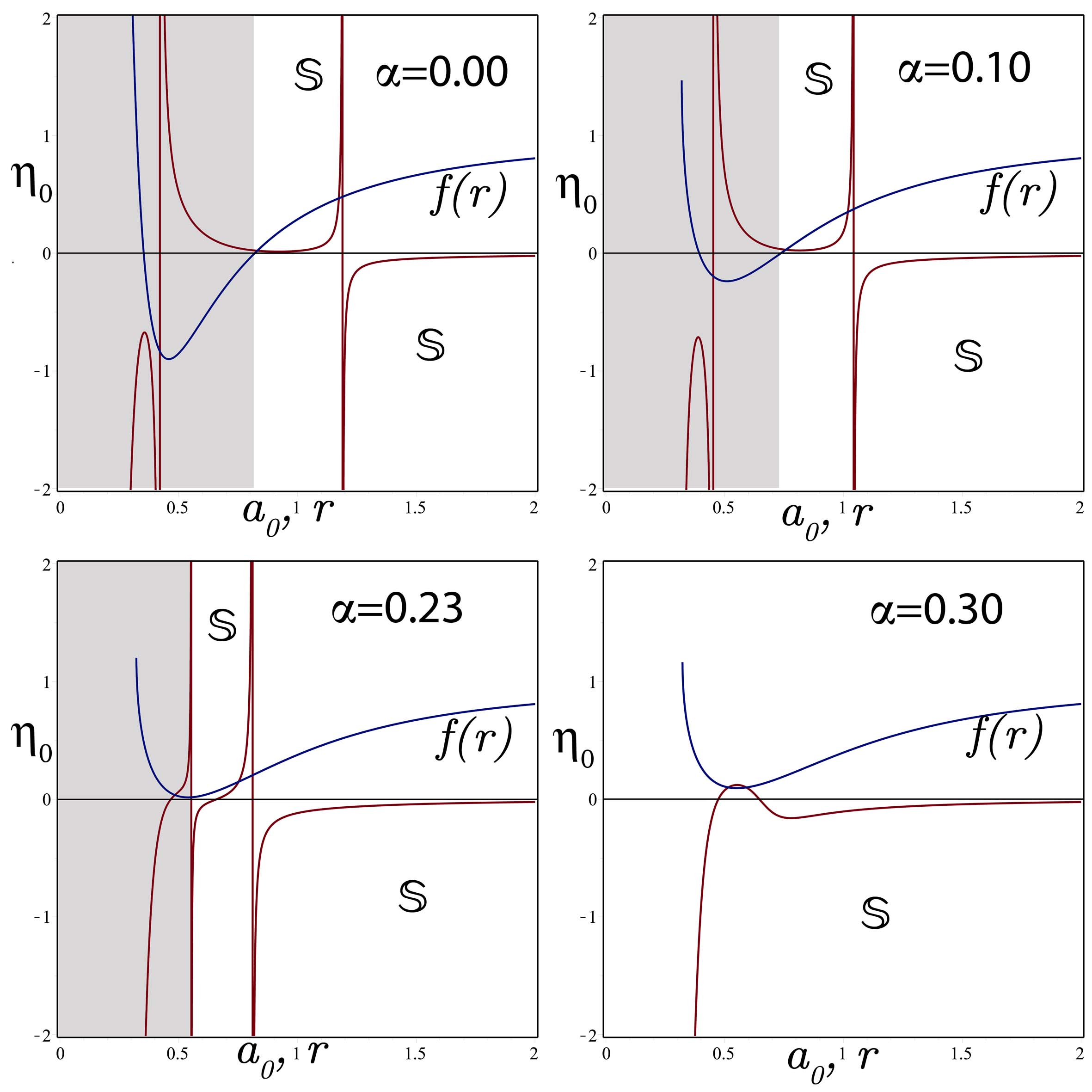}
\caption{Stability region in terms of $\protect\eta _{0}$ and radius of the
throat $a_{0}$ for MCG ($\protect\nu =1,\protect\xi _{0}=1$)$,$ $d=5,$ $M=20$%
, $Q=1$ and various values of $\protect\alpha .$ The stable region is
denoted by $\mathrm{S.}$ The metric function is also displayed in terms of $%
r $. The shaded region is for $r<r_{h}$ in which $r_{h}$ is the event
horizon.}
\end{figure}

\section{Stability of the EMGB TSW supported by an arbitrary equation of
state}

\begin{figure}[tbp]
\includegraphics[width=70mm,scale=0.7]{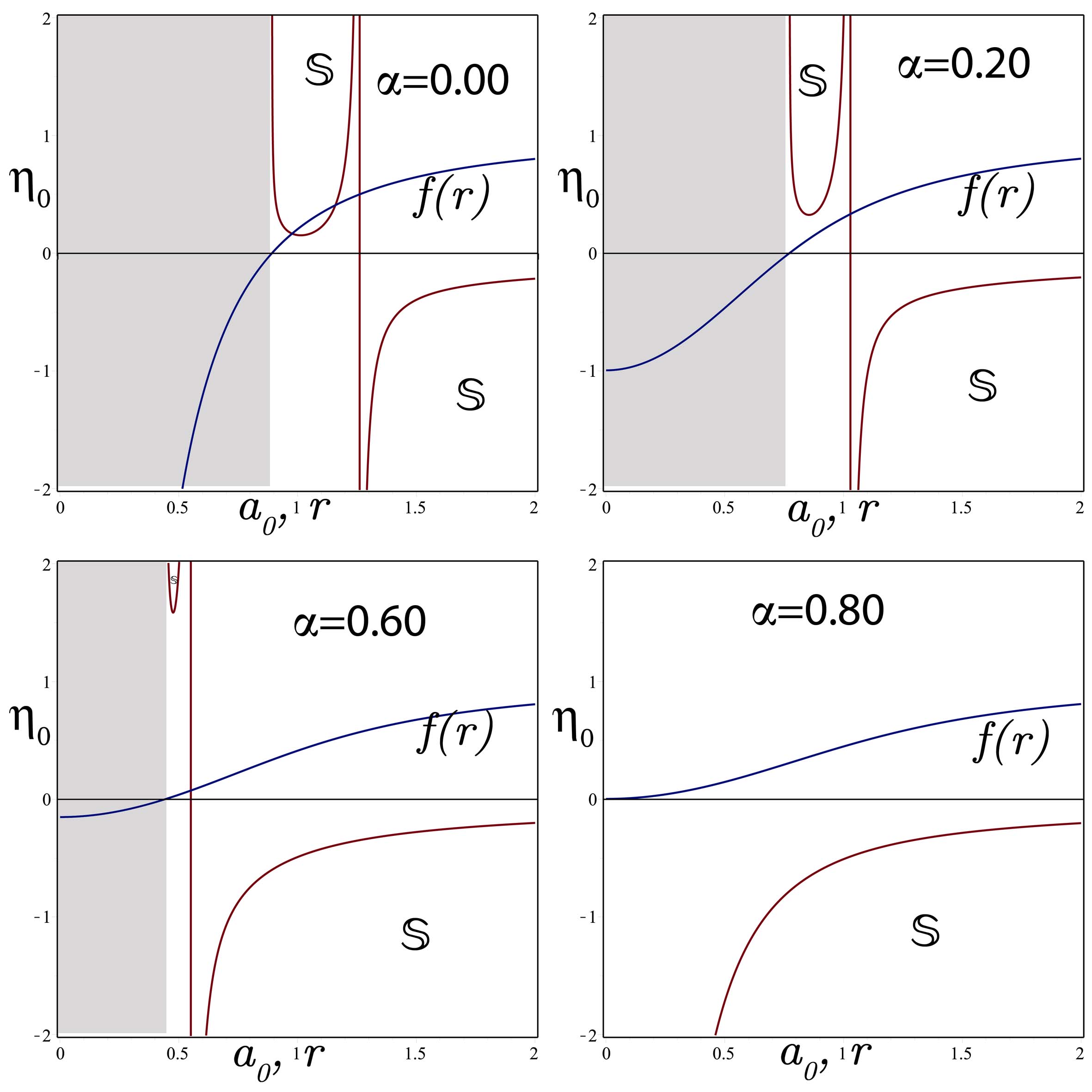}
\caption{Stability region in terms of $\protect\eta _{0}$ and radius of the
throat $a_{0}$ for logarithmic gas (LG) for $d=5,$ $M=20$, $Q=0$ and various
values of $\protect\alpha .$ The stable region is denoted by $\mathrm{S.}$
The metric function is also displayed in terms of $r$. The shaded region is
for $r<r_{h}$ in which $r_{h}$ is the event horizon.}
\end{figure}

\begin{figure}[tbp]
\includegraphics[width=70mm,scale=0.7]{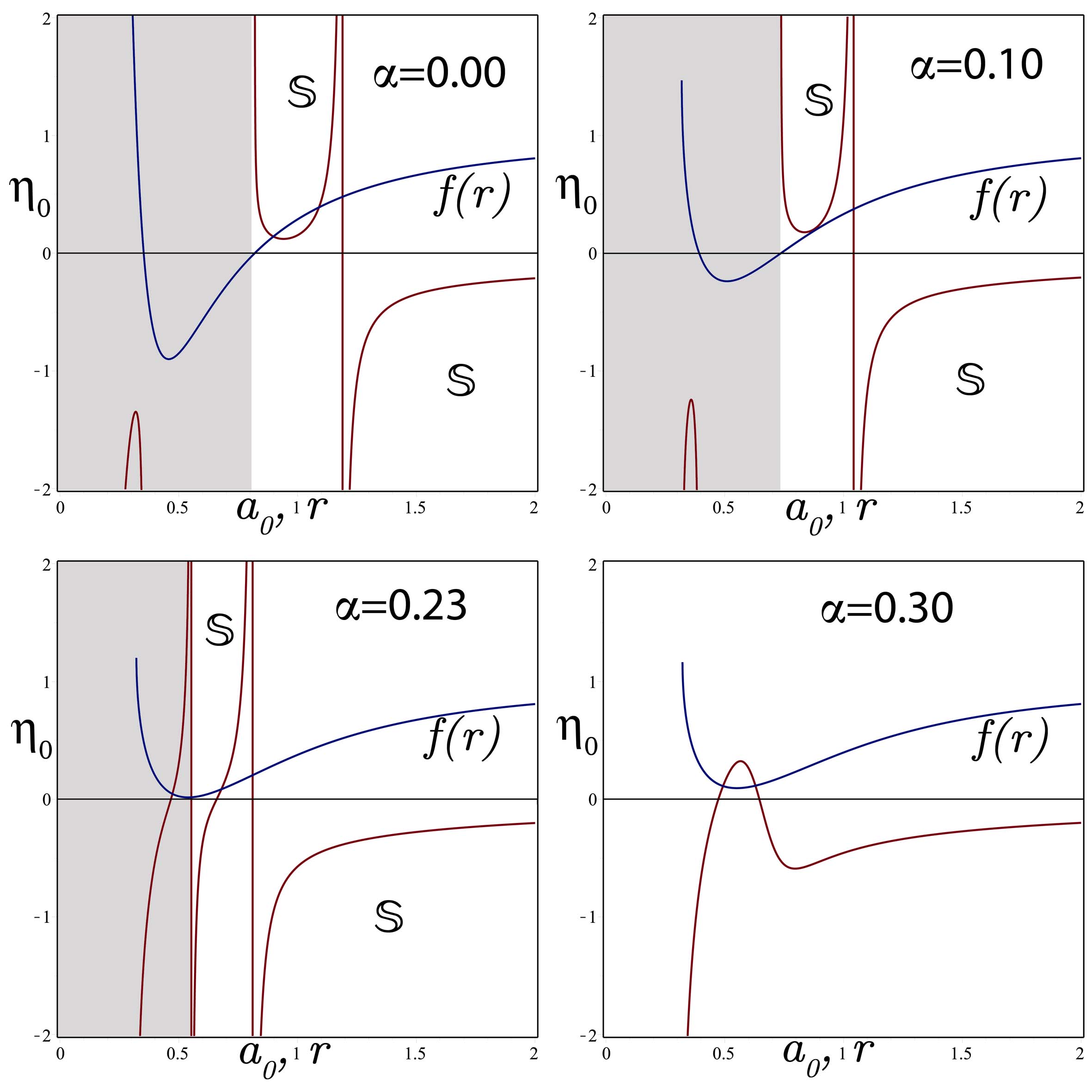}
\caption{Stability region in terms of $\protect\eta _{0}$ and radius of the
throat $a_{0}$ for LG for $d=5,$ $M=20$, $Q=1$ and various values of $%
\protect\alpha .$ The stable region is denoted by $\mathrm{S.}$ The metric
function is also displayed in terms of $r$. The shaded region is for $%
r<r_{h} $ in which $r_{h}$ is the event horizon.}
\end{figure}

In this section we study the stability of the EMGB TSW which is supported by
an arbitrary gas with the barotropic equation of state 
\begin{equation}
p=\psi \left( \sigma \right) 
\end{equation}%
in which $\psi \left( \sigma \right) $ is an arbitrary function of $\sigma .$
This covers naturally the polytropic equation of state $p\sim \sigma ^{1+%
\frac{1}{n}}$ with the index $0\leq n<\infty .$ As before, we consider the
static equilibrium configuration at $a=a_{0}$ where $\sigma _{0}$ and $p_{0}$
are given by (18) and (19). Furthermore, the equation of motion of the
throat after the perturbation is still given by (23) where $V\left( a\right) 
$ satisfies the condition (24) in which $\sigma $ in the left hand side is
the energy density after the perturbation. The form of $\sigma ,$
explicitly, depends on the form of $p=\psi \left( \sigma \right) ,$ can be
found by applying the energy conservation law (20) which is also equivalent
with%
\begin{equation}
\sigma ^{\prime }=-\frac{d-2}{a}\left( \sigma +p\right) .
\end{equation}%
Furthur, one has%
\begin{equation}
\sigma ^{\prime \prime }=-\frac{\left( d-2\right) }{a}p^{\prime }+\frac{%
\left( d-1\right) \left( d-2\right) }{a^{2}}\left( \sigma +p\right) 
\end{equation}%
in which a prime denotes derivative with respect to $a$. Having $p^{\prime
}=\psi ^{\prime }\left( \sigma \right) \sigma ^{\prime }$ the latter
equation reads%
\begin{equation}
\sigma ^{\prime \prime }=\frac{\left( d-2\right) \left( \sigma +p\right) }{%
a^{2}}\left[ \left( d-2\right) \psi ^{\prime }\left( \sigma \right) +\left(
d-1\right) \right] .
\end{equation}%
Nevertheless, using (31) and (33), one can explicitly find the form of $%
V^{\prime }\left( a\right) $ and $V^{\prime \prime }\left( a\right) $ from
(24) and show that at $a=a_{0}$, $V\left( a_{0}\right) $ and $V^{\prime
}\left( a_{0}\right) $ vanish while 
\begin{equation}
V^{\prime \prime }\left( a_{0}\right) =\frac{2\left( d-2\right) f_{0}\psi
^{\prime }\left( \sigma _{0}\right) \mathfrak{G}_{1}+\mathfrak{G}_{2}\tilde{%
\alpha}+2a_{0}^{2}\mathfrak{G}_{3}}{2a_{0}^{2}f_{0}\left[ a_{0}^{2}+2\tilde{%
\alpha}\left( 1+f_{0}\right) \right] }
\end{equation}%
in which%
\begin{multline}
\mathfrak{G}_{1}=4\tilde{\alpha}f_{0}^{2}- \\
\left( 2\tilde{\alpha}a_{0}f_{0}^{\prime }+12\tilde{\alpha}%
+2a_{0}^{2}\right) f_{0}+a_{0}f_{0}^{\prime }\left( a_{0}^{2}+2\tilde{\alpha}%
\right) ,
\end{multline}%
\begin{multline}
\mathfrak{G}_{2}=8\left( d-5\right) f_{0}^{3}+ \\
f_{0}^{2}\left[ -4a_{0}^{2}f_{0}^{\prime \prime }-4f_{0}^{\prime }\left(
d-7\right) a_{0}-24\left( d-5\right) \right] + \\
4a_{0}\left[ \left( f_{0}^{\prime \prime }-\frac{f_{0}^{\prime 2}}{2}\right)
a_{0}+f_{0}^{\prime }\left( d-7\right) \right] f_{0}-2a_{0}^{2}f_{0}^{\prime
2},
\end{multline}%
\begin{equation}
\mathfrak{G}_{3}=a_{0}^{2}\left( f_{0}f_{0}^{\prime \prime }-\frac{%
f_{0}^{\prime 2}}{2}\right) +\left( f_{0}f_{0}^{\prime
}a_{0}-2f_{0}^{2}\right) \left( d-3\right) .
\end{equation}%
We note that $\psi ^{\prime }\left( \sigma _{0}\right) \left( =\frac{%
p_{0}^{\prime }}{\sigma _{0}^{\prime }}\right) =\left. \frac{d\psi }{d\sigma 
}\right\vert _{\sigma =\sigma _{0}}$ while the other functions are
calculated at $a=a_{0}.$ Depending on the form of $\psi $ we face different
TSW. For instance setting $\frac{d\psi }{d\sigma }=\eta _{0}=$constant
reduces to a linear gas supporting TSW with 
\begin{equation}
\psi =\eta _{0}\sigma +C
\end{equation}%
where $C$ is a constant. Imposing $p\left( a=a_{0}\right) =p_{0}$ and $%
\sigma \left( a=a_{0}\right) =\sigma _{0}$ leads to $C=p_{0}-\eta _{0}\sigma
_{0}$ and therefore 
\begin{equation}
\psi =\eta _{0}\left( \sigma -\sigma _{0}\right) +p_{0}
\end{equation}%
which is the case studied in \cite{26}. Another interesting case is given by 
$\frac{d\psi }{d\sigma }=-\frac{\eta _{0}}{\sigma ^{2}}$ giving%
\begin{equation}
\psi =\frac{\eta _{0}}{\sigma }+C
\end{equation}%
in which $C$ is an integration constant. Again imposing $p\left(
a=a_{0}\right) =p_{0}$ and $\sigma \left( a=a_{0}\right) =\sigma _{0}$
dictates that $C=p_{0}-\frac{\eta _{0}}{\sigma _{0}}$ and therefore%
\begin{equation}
\psi =\eta _{0}\left( \frac{1}{\sigma }-\frac{1}{\sigma _{0}}\right) +p_{0}.
\end{equation}%
Setting $p_{0}-\frac{\eta _{0}}{\sigma _{0}}=0$ or $\eta _{0}=p_{0}\sigma
_{0}$ implies the well known CG which we have studied in the previous
chapter i.e., 
\begin{equation}
\psi =p_{0}\frac{\sigma _{0}}{\sigma }.
\end{equation}%
Another important state that has been considered recently is the modified
generalized Chaplygin gas MGCG obtained by setting 
\begin{eqnarray}
\frac{d\psi }{d\sigma } &=&\xi _{0}+\frac{\nu \eta _{0}}{\sigma ^{\nu +1}} \\
&&\left( \xi _{0}=\text{constant}\right) 
\end{eqnarray}%
which implies%
\begin{equation}
\psi =\xi _{0}\sigma -\frac{\eta _{0}}{\sigma ^{\nu }}+C.
\end{equation}%
Applying $p\left( a=a_{0}\right) =p_{0}$ and $\sigma \left( a=a_{0}\right)
=\sigma _{0}$ yields $C=p_{0}+\frac{\eta _{0}}{\sigma _{0}^{\nu }}-\xi
_{0}\sigma _{0}$ and consequently%
\begin{equation}
\psi =\xi _{0}\left( \sigma -\sigma _{0}\right) -\eta _{0}\left( \frac{1}{%
\sigma ^{\nu }}-\frac{1}{\sigma _{0}^{\nu }}\right) +p_{0}.
\end{equation}%
Setting $C=0$ or $\eta _{0}=\sigma _{0}^{\nu }\left( \xi _{0}\sigma
_{0}-p_{0}\right) $ simplifies the latter equation as%
\begin{equation}
\psi =\xi _{0}\sigma -\frac{\eta _{0}}{\sigma ^{\nu }}
\end{equation}%
which has been studied in \cite{27}. Fig. 2 depicts the effect of GB
parameter on the stability regions of the CG model of TSWH in pure GB
gravity (i.e., $Q=0$). It is observed that increasing the value of the GB
parameter decreases the stability areas. Fig. 2 displays stability regions
as Fig. 1 but with $Q=1.$ Almost the same effect of GB parameter is seen in
this case too. We note from the standard CG model that $0<\eta _{0}$ while
the figures are plotted for $-2<\eta _{0}\leq 2.$ What we are referring to
as the stability region should be understood in this interval.

Figs. 4 and 5 are plots of stability regions for TSW in EGB ($Q=0$) and EMGB
($Q=1$) supported by MCG ($\xi _{0}\neq 0,\eta _{0}\neq 0,\nu =1$). \ Fig. 4
should be compared with Fig.\ 2 and Fig. 5 should be compared with Fig.\ 3
to see the change of the stability of the TSWH in EGB and EMGB bulk due to
MCG instead of CG. We observe that effects of MCG becomes more significant
for the regions of stability $r<r_{h}$ and for the cases which admits no
horizon.

\subsection{A Logarithmic model of gas supporting the TSW in EMGB gravity}

As one can see from Eq. (34), in $V^{\prime \prime }\left( a_{0}\right) $
only $\psi ^{\prime }\left( \sigma _{0}\right) $ appears. In the case of GCG
i.e. $\psi =-\frac{\eta _{0}}{\sigma ^{\nu }}$ with $0<\eta _{0}$ and $0<\nu
\leq 1,$ $\psi ^{\prime }\left( \sigma \right) =\frac{\nu \eta _{0}}{\sigma
^{\nu +1}}.$ We note that the case $\nu =0$ is excluded, for this reason
separately we consider the case $\nu =0$ briefly here. When $\nu =0,$ $\psi
^{\prime }\left( \sigma \right) =\frac{\eta _{0}}{\sigma }$ which implies $%
\psi =\eta _{0}\ln \left\vert \sigma \right\vert +C.$ In Figs. 6 and 7 we
plot the stability regions of the TSW supported by the Logarithmic state
equation in EGB and EMGB bulk metrics respectively.

\section{Conclusion}

In conclusion, for a GCG obeying the equation of state $p=\left( \frac{%
\sigma _{0}}{\sigma }\right) ^{\nu }p_{0}$, we have found stable regions
within physically acceptable range of parameters in EMGB gravity. The role
of GB parameter $\alpha $ in the formation of stable TSW is investigated. It
is found that formation of stable regions is highly dependent on the value
of $\alpha $ as depicted in our numerical plots. The energy-density,
however, turns out to be negative to suppress such a TSW as a prominent
candidate. Besides, a general equation of state is considered in the form $%
p=\psi \left( \sigma \right) $ which reproduces all known particular cases.
It is found that depending on tuning of the parameters stable regions expand
/ shrink accordingly. Unfortunately in all cases tested one had to be
satisfied with a negative energy density as the supporting agent for the TSW
in EMGB theory. Finally we wish to comment that in addition to the classical
role played by wormholes their possible quantum roles within the context of
"firewalls paradox" has recently been highlighted \cite{28}. It is
speculated that the emitted Hawking particles are entangled through
wormholes to the inner-horizon particles of a black hole \cite{29}. Once
justified, the subject of wormholes will turn into a hot topic to transcend
classical boundaries to occupy a significant role even in quantum gravity.

\end{document}